\documentclass[floats,aps,epsf,amsfonts,nofootinbib,showpacs]{revtex4-1}
\usepackage[all]{xy}
\usepackage{amsmath,amsthm}
\usepackage[dvips]{graphicx}
\usepackage{comment} 
\usepackage{array}
\usepackage{latexsym,amssymb}
\usepackage{footmisc}

\newcommand{\beq}{\begin{equation}}
\newcommand{\eeq}{\end{equation}}

\newcommand{\vf}{\varphi}

\renewcommand{\th}{\theta}
\renewcommand{\BibitemShut}[1]{}

\begin{document}

\title{Self-force from reconstructed metric perturbations: numerical implementation in Schwarzschild spacetime
}
\author{Cesar Merlin and Abhay G.\ Shah}
\affiliation{Mathematical Sciences, University of Southampton, Southampton, SO17 1BJ, United Kingdom}

\begin{abstract}
We present a first numerical implementation of a new scheme by Pound {\it et al.}\ \citep{BMP1} that enables the calculation of the gravitational self-force in Kerr spacetime from a reconstructed metric-perturbation in a radiation gauge. The numerical task of the metric reconstruction essentially reduces to solving the fully separable Teukolsky equation, rather than having to tackle the linearized Einstein's equations themselves in the Lorenz Gauge, which are not separable in Kerr. The method offers significant computational saving compared to existing methods in the Lorenz gauge, and we expect it to become a main workhorse for precision self-force calculations in the future. Here we implement the method for circular orbits on a Schwarzschild background, in order to illustrate its efficacy and accuracy. We use two independent methods for solving the Teukolsky equation, one based on a direct numerical integration, and the other on the analytical approach of Mano, Suzuki, and Takasugi. The relative accuracy of the output self-force is at least $10^{-7}$ using the first method, and at least $10^{-9}$ using the second; the two methods agree to within the error bars of the first. We comment on the relation to a related approach by Shah {\it et al}. \citep{friedman2}, and discuss foreseeable applications to more generic orbits in Kerr spacetime. 
\end{abstract}
\date{\today}
\maketitle
\section{Introduction}
The relativistic two-body problem can be tackled using black hole perturbation theory in the extreme mass ratio inspiral (EMRI) regime, in which one of the components is much larger than its companion. The smaller object experiences a self-force (SF) due to the interaction with its own gravitational field. We may identify two pieces of the SF, the conservative and the dissipative. The dissipative piece of the SF is responsible for the loss of energy and angular momentum of the orbiting bodies which are radiated away as gravitational waves.
The conservative piece of the SF modifies the positional elements of the orbit; for example, it is responsible for the shift in orbital precession \citep{pound1,pound2}.

The equation of motion for a small mass moving in a curved spacetime was originally formulated by Mino, Sasaki and Tanaka \citep{MiSa}, and independently by Quinn and Wald \citep{QuWa} ---the resulting equation is usually referred to simply as the MiSaTaQuWa equation. The MiSaTaQuWa SF was formulated in the Lorenz gauge where the field equations become hyperbolic, which makes them suitable to solve numerically, and the singularity of the particle's representation is locally-isotropic. The behaviour of the SF under a gauge transformation was studied by Barack and Ori in \citep{barack1}, where they showed how to compute it in any gauge related to Lorenz's via a sufficiently regular transformation.

Current calculations of the SF usually rely on numerical solutions of the linearized Einstein's equations in the Lorenz gauge \cite{Barack:2005nr}. In Schwarzschild this involves solving ten coupled differential equations for the tensorial-harmonic components of the perturbation. With the metric perturbations as an input one may obtain the SF using the mode-sum method \citep{leorD61} or the puncture method \citep{barackD76, VegDet}. On Kerr spacetime the tensorial field equations in the Lorenz gauge (LG) are not separable and one has to deal with a system of partial differential equations. This has been a motivation to work in time-domain implementations \cite{dolan1, dolan2, dolan3, vega, canizares, wardell2011} of MiSaTaQuWa formula with a puncture, but the numerical evolution in this scheme takes considerably more time than frequency-domain implementations. 

The numerical treatment of black hole perturbations in Kerr spacetime becomes much simpler in a radiation gauge (RG) where one implements the Chrzanowski-Cohen-Kegeles-Wald (CCK) \cite{chrza, cohen, waldrec} formalism to reconstruct the metric perturbations from the perturbed spin-$\pm 2$ Weyl scalars, which are solutions to the separable Teukolsky equation, using an intermediate Hertz potential. Recently, Pound {\it et al}. \citep{BMP1} provided the necessary framework to regularize the force calculated from the RG reconstructed perturbations. A numerical prescription to calculate the SF using the metric perturbations (MP) in the Outgoing Radiation Gauge (ORG) was presented by Shah {\it et al}.\ \citep{friedman2} for a particle in a circular orbit around a Schwarzschild black hole. This work assumed that the LG mode-sum would remain valid in the RG and the authors found numerical confirmation that some of the LG regularization parameters could regularize the force in the RG. An important result from \citep{friedman2} is the computation of gauge-invariant quantities ($H\equiv \frac{1}{2}h_{\alpha\beta}u^\alpha u^\beta$, where $h_{\alpha\beta}$ is the reconstructed MP in the radiation gauge and $u^\alpha$ is the four-velocity of the particle) from RG modes in Schwarzschild and a comparison with the LG values. An extension to a Kerr background for the CCK metric reconstruction has lead to a successful calculation of $H$ \citep{friedman3}. Pound {\it et al}.\ \citep{BMP1} identified three categories of radiation gauges (``full-'',``half-'', and ``no-'' string gauges) according to the singular structure of the  MP. In the full-string RG the singularity extends along a radial null direction of the spacetime and through the particle at each time; in the half-string gauges the singularity is confined either inside or outside the 2-sphere intersecting the particle (at a given time and radius); for the no string gauges the MP has no singularity but it exhibits a discontinuity at the 2-sphere containing the particle (see also \citep{CanSop}). In none of the above cases is the singularity confined to the location of the particle, in contrast to the LG. As a main practical result of these analyses Pound {\it et al}.\ found the averaged version of the mode-sum for the no string gauges that we implement in this work --- they also found non-trivial modifications to the standard LG mode-sum formula for the half string gauges. Unlike \citep{friedman2}, in this work we re-expand the expression of the retarded force calculated from the RG metric perturbations in terms of the usual scalar spherical harmonics for which the original mode-sum was derived.

The structure of this paper is as follows. In Sec. \ref{forma} we present a short review of the formalism required in our implementation including the mode-sum formula using the RG modes, the metric reconstruction from solutions to Teukolsky equation and the inclusion of low multipoles using the analytical expressions obtained by Barack and Lousto \citep{Barack:2005nr}. The algorithm of the numerical implementation to calculate the radial and temporal components of the SF are given in Sec. \ref{num} together with a short review of the Mano-Susuki-Takasugi method. The numerical results are given in Sec. \ref{results} where we show that our implementation is consistent with the existing literature (energy flux and $t$-component of the SF). For completeness we include full expressions for the static modes, Sasaki-Nakamura equation, Teukolsky sources and retarded force in terms of spin-weighted harmonics in the appendix.

In this work the metric signature is $(-,+,+,+)$, Greek letters are used for spacetime Boyer--Lindquist coordinates $(t,r,\theta,\varphi)$ indices, and we work in standard geometrized units (with $c=G=1$). We will denote the complex conjugated of a quantity or operator by $\bar{}$ on top of it. Bold indices correspond to projections with respect of the Kinnersly tetrad $(\ell ,n,m,\bar m)$.

\section{Review of the formalism}\label{forma}
The gravitational force acting on a particle of mass $\mathsf{m}$ due to a smooth external perturbation $h_{\alpha\beta}$ (at the particle's location $x^\alpha=x^\alpha_0$) is given according to \citep{Barack:2009ux} by
\begin{equation}\label{force}
F^\alpha (x_0)=-\lim_{x\rightarrow x_0}\mathsf{m}\left(g^{\alpha\beta}+u^\alpha u^\beta\right)\left[ \nabla_\mu h_{\nu\beta}(x)-\frac{1}{2}\nabla_\beta h_{\mu\nu}(x)\right]u^\mu u^\nu,
\end{equation}
where $g_{\alpha\beta}$ is the background metric (latter we will specialize to the Schwarzschild metric), $\nabla_\alpha$ is the metric compatible covariant derivative and the four-velocity of the particle $u^\alpha\equiv dx_0^\alpha(\tau)/d\tau$, $\tau$ being the proper time, has to be extended off the world-line to make sense of the limit. In principle one can also extend the covariant derivatives and $g^{\alpha\beta}$---since the resulting force only has support on the worldline of the particle--- or equivalently leave them as fields and take the limit consistently. The value of the force is independent of the extension chosen.

\subsection{Mode-Sum regularization}
\label{modesumform}
Consider a small mass in geodesic motion around a Kerr black hole. The perturbed metric due to the presence of the small mass $g+h$ diverges at the location of the particle $x_0$. Detweiler and Whiting showed that the full retarded metric perturbation admits a decomposition into certain locally defined {\it singular} piece $h^S_{\alpha\beta}$ and a smooth {\it regular} field $h^{\rm R}_{\alpha\beta}$ \citep{DetWhi}, namely
\begin{align}
h^\textrm{full}_{\alpha\beta}=h^\textrm{S}_{\alpha\beta}+h^{\rm R}_{\alpha\beta},
\end{align}
where $h^S_{\alpha\beta}$ is chosen near the location of the particle to cancel the singular part of $h^{full}_{\alpha\beta}$ while not contributing to the SF. Each component of the SF can be obtained by subtracting the singular part of the force from the {\it full} (or {\it retarded}) value
\begin{equation}
F^\alpha_\textrm{self}(x_0)=\lim _{x\rightarrow x_0} \left[F^\alpha_\textrm{full}(x)-F^\alpha_\textrm{S}(x)\right],
\end{equation}
where the fields $F^\alpha_{full}(x)$ and $F^\alpha_S(x)$ satisfy Eq.\ \eqref{force} with $h_{\mu\nu}^{full}$ and $h_{\mu\nu}^S$ respectively.

A practical way to implement Eq.\ \eqref{force} and obtain the SF is given by the mode-sum regularization procedure. The fields $F^\alpha_\textrm{full}(x)$ and $F^\alpha_\textrm{S}(x)$ can be expanded in spherical harmonics $Y_{\ell m}(\theta,\varphi)$ on the surface $t,r=cons$.\ (ignoring the vectorial nature of the SF and treating each of its Boyer-Linquist components as a scalar function; see \citep{haaspoisson} for a more sophisticated covariant approach). By subtracting the desired $\ell$-mode contributions (summed over all possible values of $m$) for the full and the singular pieces of the force \citep{Barack:2001bw} we obtain
\begin{align}
\label{Full-S}
F^\alpha_\textrm{self}(x_0)=\lim_{x\rightarrow x_0}\sum^{\infty}_{\ell=0}\left[F^{\alpha\ell}_\textrm{full}(x)-F^{\alpha\ell}_\textrm{S}(x)\right].
\end{align}

The quantities $F^{\alpha}_\textrm{full}(x)$ and $F^{\alpha}_\textrm{S}(x)$ diverge at $x\rightarrow x_0$ (since the {\it full} and {\it singular} MP diverge at $x_0$). However, each of the individual $\ell$-modes $F^{\alpha\ell}_\textrm{full}(x)$ and $F^{\alpha\ell}_\textrm{S}(x)$ are finite. The difference $F^{\alpha\ell}_\textrm{full}(x)-F^{\alpha\ell}_\textrm{S}(x)$ correspond to the $\ell$-mode of $F^\alpha_\textrm{self}(x)$ which is smooth everywhere.

It is known that in the Lorenz gauge $\lim_{x\to x_0}F^{\alpha\ell}_\textrm{S}(x)$ has the large-$\ell$ expansion $\lim_{x\to x_0}F^{\alpha\ell}_\textrm{S}(x)=A^\alpha L+B^\alpha +C^\alpha /L+...$ \citep{Barack:2001bw}, with $L \equiv \ell +1/2$. The coefficients $A^\alpha$, $B^\alpha$ and $C^\alpha$ are the $\ell$-independent regularization parameters for each component of the SF. The sum in \eqref{Full-S} converges faster than any power of $1/\ell$ (recall $F^\alpha_\textrm{full}(x)-F^\alpha_\textrm{S}(x)$ is smooth). We expect that both the full and singular pieces share the same large-$\ell$ power expansion with the same coefficients. We can then express Eq.\ \eqref{Full-S} as a difference of two convergent sums, in the form
\begin{align}
F^\alpha_\textrm{self}(x_0)=&\sum^\infty_{\ell =0}\left[F_{{\rm full}\pm}^{\alpha\ell}(x_0)\mp A^\alpha L-B^\alpha-C^\alpha/L\right] \nonumber \\
 &-\sum^\infty_{\ell=0}\left[F_{S\pm}^{\alpha\ell}(x_0)\mp A^\alpha L-B^\alpha-C^\alpha/L\right], 
\end{align}
where we first take the limits $t\to t_0$, $\theta\to \theta_0$, $\varphi\to\varphi_0$ and the sign $\pm$ depends on the side we approach the value of $r_0$ (the sum $F^{\alpha\ell}_{{\rm full}\pm}(x_0)\mp A^\alpha L $ is direction independent). The individual terms of the sums go as $\sim 1/\ell^2$ and the sequence of partial sums converges as $\sim 1/\ell$. We arrive at 
\begin{align}
\label{modesum}
F^\alpha_\textrm{self}(x_0)=\sum^\infty_{\ell=0}\left(F_{{\rm full} \pm}^{\alpha\ell}(x_0)\mp A^\alpha L-B^\alpha-C^\alpha/L\right)-D^\alpha,
\end{align}
with
\begin{align}
D^\alpha\equiv\sum^\infty_{\ell=0}\left(F_{S\pm}^{\alpha\ell}(x_0)\mp A^\alpha L-B^\alpha -C^\alpha/L\right).
\end{align}
In the LG the analytical form of the regularization parameters $A^\alpha$ and $B^\alpha$ is well known in Kerr \citep{barackelgr} with $C^\alpha \equiv D^\alpha\equiv 0$. The values of the regularization parameters remain invariant under gauge transformations from LG that are sufficiently regular \citep{barack1}. 

However, as shown in \citep{BMP1}, the gauge transformation vector that goes from Lorenz to Radiation gauge is either singular along a radial null direction (at each time) ---in the best case scenario this singularity is present only in half of the spacetime either in $r>r_0$ or $r<r_0$ at constant $t$--- or it is discontinuous at the 2-sphere of radius $r=r_0$.

For a discontinuous RG ---the one that transforms to Lorenz via a discontinuous gauge vector--- a two-sided average  mode-sum formula still holds true due to the parity regularity of the transformation vector \citep{BMP1}:  
\begin{align}
\label{modesumave}
F^\alpha_\textrm{self}=\sum_{\ell}\left[\frac{1}{2}\left(F^\alpha \right)^\ell_+ +\frac{1}{2}\left(F^\alpha\right)^\ell_- -B^\alpha-C^\alpha/L\right]-D^\alpha,
\end{align}
where $\left(F^\alpha \right)^\ell_\pm$ is short hand for $\lim_{x\rightarrow \pm x_0}F_{{\rm full}\pm}^{\alpha\ell}(x)$. The regularization parameters $B^\alpha$, $C^\alpha$ and $D^\alpha$ take the standard LG values for the chosen extension. 

The regularization parameters $A^r$ and $B^r$ for circular orbits in Schwarzschild are given analytically (the Schwarzschild coordinate components of $u^\alpha$ are extended as constant fields away from $x_0$ and the metric related quantities take their field value \footnote{See \citep{baracksf} for a full derivation.}) by
\begin{align}
\label{regpar}
A^r_\pm=\mp\frac{\mathsf{m}^2{\cal E}}{r_0^2f_0^2\tilde V},\qquad B^r=-\frac{\mathsf{m}^2}{r_0^2}\frac{2{\cal E}^2\hat K(\omega)-{\cal E}^2\hat E(\omega)}{\pi f_0^2 \tilde V^{3/2}},
\end{align}

with $\tilde V=1+\frac{{\cal L}^2}{r_0^2}$, $\omega={\cal L}^2/({\cal L}^2+r_0^2)$ and $f_0\equiv 1-2M/r_0$. We have also used the orbital parameters defined as
\begin{align}
\label{orbpar}
\Omega\equiv\frac{u^\varphi}{u^t}=\sqrt{\frac{M}{r_0^3}},\qquad{\cal E}\equiv \frac{r_0-2M}{\sqrt{r_0^2-3Mr_0}},\qquad {\cal L}\equiv\sqrt{\frac{r_0^2 M}{r_0-3M}},
\end{align}
which correspond to the orbital frequency, the specific energy and angular momentum of the particle around a circular orbit of radius $r_0$. The functions $\hat K(\omega)$ and $\hat E(\omega)$ are the complete elliptic integrals of first and second kind respectively. The sign $\pm$ in $A^r$ refers to the sided radial limit once again. In virtue of using the two-sided average version of the mode-sum the contributions from $A^r$ to the mode-sum formula will cancel [see Eq.\ \eqref{modesumave}]. Let us stress that the analytical expression for $B^r$ is extension dependent and only by consistently using the same extension throughout the calculation the mode-sum method will give the correct value of the SF for a given gauge.

\subsection{Newman-Penrose Formalism and metric reconstruction}
In Schwarzschild spacetime, the Kinnersley tetrad in Boyer-Linquist coordinates is given by
\begin{align}
\ell^\alpha=\left(\frac{1}{f(r)},1,0,0\right),\quad n^\alpha=\frac{1}{2}\left(1,-f(r),0,0\right),\quad m^\alpha=\frac{1}{\sqrt{2} r}\left(0,0,1,\frac{i}{\sin\th}\right),
\end{align}
where $f(r)\equiv 1-2M/r$. We will omit the explicit functional dependence of $f(r)$ to simplify the notation. The corresponding directional derivatives are denoted by $\boldsymbol{D}\equiv \ell^\alpha \nabla_\alpha$, $\boldsymbol{\Delta }\equiv n^\alpha \nabla_\alpha$, $\boldsymbol{\delta }\equiv m^\alpha \nabla_\alpha$. The non-zero spin coefficients are
\begin{align}
\varrho=-\frac{1}{r}, \quad \beta=-\alpha=\frac{\cot\th}{2\sqrt{2} r},\quad\gamma=\frac{M}{2r^2},\quad\mu=-\frac{1}{2r}f.
\end{align}
Teukolsky equation \citep{Teuk} describes perturbations of a Kerr space-time with source $T_s$ [given below in Eq.\ \eqref{kerrsources}]. Let us specialize to the  Schwarzschild case by setting $a=0$. For a particular value of spin $s$ the Bardeen-Press equation (Teukolsky equation with $a = 0$) is given by
\begin{align}
\label{kerrteuk}
\frac{r^2}{f} \frac{\partial^2\psi_s}{\partial t^2}-\left(\frac{1}{\sin^2\theta}\right)\frac{\partial^2\psi_s}{\partial\varphi^2}-(r^2 f)^{-s}\frac{\partial}{\partial r}\left[(r^2f)^{s+1}\frac{\partial \psi_s}{\partial r}\right]
-\frac{1}{\sin\theta}\frac{\partial}{\partial\theta}\left(\sin\theta \frac{\partial\psi_s}{\partial\theta}\right)\nonumber \\-2s\frac{i\cos\theta}{\sin^2\theta}\frac{\partial\psi_s}{\partial\varphi} 
-2s\frac{(-r+3M)}{f}\frac{\partial\psi_s}{\partial t}
+(s^2\cot^2\theta -s)\psi_s =-4\pi r^2 T_s,
\end{align}
where $\psi_s$ in the frequency-domain separates as follows:
\begin{align}
\psi_s  =& e^{-i\omega t}R_s(r)\,_sY_{\ell m}(\th,\vf), \quad \textrm{with}\qquad \omega\equiv m \Omega.
\end{align}
The eigenfunctions of the angular part of Eq.\ \eqref{kerrteuk} are the spin-weighted spherical harmonics $_sY_{\ell m}(\th,\vf)$ and the function $R_s(r)$ is solution of the radial part of Eq.\ \eqref{kerrteuk}.

The relevant gravitational sources ($T_{2}$ for $\psi_{s=2}\equiv\psi_0$ and $T_{-2}$ for $\psi_{-2}\equiv \varrho^{-4}\psi_4$ respectively) are given by
\begin{subequations}\label{kerrsources}
\begin{align}
T_{-2}=&2 \varrho^{-4}\left\{ (\boldsymbol{\Delta } +2\gamma +5\mu)\left[ (\boldsymbol{\bar\delta}+2\alpha )T_{\boldsymbol{24}}-(\boldsymbol{\Delta }+\mu )T_{\boldsymbol{44}}\right]  +(\boldsymbol{\bar\delta} +2\alpha )\left[ (\boldsymbol{\Delta } +2\gamma +2\mu ) T_{\boldsymbol{24}}-\boldsymbol{\bar\delta}T_{\boldsymbol{22}} \right]\right\}, \label{kerrsource-2}\\
T_{2}=&2\left\{ (\boldsymbol{\delta } -2\beta )\left[ (\boldsymbol{D} -2\varrho )T_{\boldsymbol{13}}-\boldsymbol{\delta }T_{\boldsymbol{11}}\right]  +(\boldsymbol{D} -5\varrho)\left[ (\boldsymbol{\delta} -2\beta ) T_{\boldsymbol{13}}-(\boldsymbol{D}-\varrho )T_{\boldsymbol{33}} \right]\right\},\label{kerrsource}
\end{align}
\end{subequations}
where the projections of the stress-energy tensor are given by:%!along the Kinnersly tetrad: 
\begin{equation}
T_{\boldsymbol{ab}}\equiv e^\alpha _{\boldsymbol{a}} e^\beta _{\boldsymbol{b}} T_{\alpha\beta},
\end{equation}
with $e^{\alpha}_{\boldsymbol{a}}=(\ell^\alpha,n^\alpha, m^\alpha,\bar m^\alpha)$ and $T^{\alpha\beta}=\frac{\mathsf{m}}{u^t r_0^2}u^\alpha u^\beta \delta(r-r_0)\delta(\theta-\theta_0)\delta(\vf-\vf_0)$. The explicit expressions for $T_{\pm2}$ can be found in the Appendix \ref{IRGvsORG}.
All the components of the metric perturbation tensor are recovered by applying a differential operator on a scalar quantity $\Psi$. This {\it Hertz Potential} $\Psi$ is also solution to the homogeneous Teukolsky equation with opposite spin as the Weyl scalar from which it is constructed. The relevant operators were given by Chrzanowzki \citep{chrza} and Cohen-Kegeles \citep{cohen}
\begin{align}
\label{hORG}
h_{\alpha\beta}^{\rm ORG}=& -\varrho^{-4}\left\{n_\alpha n_\beta\left(\boldsymbol{\bar\delta}-2\alpha \right)\left(\boldsymbol{\bar\delta}-4\alpha\right)+\bar m_\alpha\bar m_\beta\left(\boldsymbol{\Delta}+5\mu-2\gamma \right)\left(\Delta +\mu -4\gamma\right)\right. \nonumber \\
 &\left. -n_{(\alpha}\bar m_{\beta)} \left[ \left(\boldsymbol{\bar\delta} -2\alpha\right)\left(\boldsymbol\Delta+\mu-4\gamma\right)+\left(\boldsymbol{\Delta} +4\mu -4\gamma\right)\left(\boldsymbol{\bar\delta} -4\alpha\right)\right]\right\}\Psi^{\rm ORG} \pm {\rm c.c.},\\
\label{hIRG}
h_{\alpha\beta}^{\rm IRG}=& \left\{-\ell_\alpha \ell_\beta\left(\boldsymbol{\delta}+2\beta \right)\left(\boldsymbol{\delta}+4\beta\right)- m_\alpha m_\beta\left(\boldsymbol{D}-\varrho\right)\left(\boldsymbol{D} +3\varrho \right)\right. \nonumber \\
 &\left. -\ell_{(\alpha} m_{\beta)} \left[ \boldsymbol{D} \left(\boldsymbol\delta+4\beta\right)+\left(\boldsymbol{\delta} +4\beta \right)\left(\boldsymbol{D} +3\rho\right)\right]\right\}\Psi^{\rm IRG} \pm {\rm c.c.},
\end{align}
where the sign $\pm$ corresponds to the state of polarization and c.c.\ stands for the complex conjugated terms.
Ori \citep{oriCCK} showed that $\Psi$ is unique for a specific gauge. To recover the correct MP using the CCK reconstruction, $\Psi$ must satisfy certain fourth order differential equation with the Weyl curvature scalars ($\psi_0$ or $\rho^{-4}\psi_4$) as source: 

\begin{subequations}\label{hertz}
\begin{align}
\rho ^{-4}\psi_4 &= \frac{r^4 f^2}{32} \tilde{\boldsymbol{D}}^4\left(r^4 f^2\bar\Psi^{\rm ORG}\right) ,\\
\psi_0 &=\frac{1}{8}\left[\eth ^4\bar\Psi^{\rm ORG} +12M\partial_t\Psi^{\rm ORG} \right],\label{hertz0}\\
\psi_0 &=\frac{1}{2}\boldsymbol{D}^4\bar\Psi^{\rm IRG}, \\
\rho^{-4}\psi_4 &=\frac{1}{8}\left[\tilde\eth^4\bar\Psi^{\rm IRG} -12M\partial_t\Psi^{\rm IRG}\right],\label{hertz4}
\end{align}
\end{subequations}
where we have used $\tilde D\equiv -\frac{1}{f}\partial_t +\partial_r$ for Schwarzschild. The operators that lower or raise the spin-weight of the angular functions $_sY_{\ell m}(\th,\vf)$ are given by
\begin{align}
\eth\eta=&-(\partial_\th +i\csc\th\partial_\vf-s\cot\th)\eta =-\sqrt{2}r\left(\boldsymbol{\delta}-2s\beta\right)\eta,\nonumber\\
\bar\eth\eta=&-(\partial_\th -i\csc\th\partial_\vf+s\cot\th)\eta =-\sqrt{2}r\left(\boldsymbol{\bar\delta}+2s\beta\right)\eta,
\end{align}
with the useful identities
\begin{align}
\eth {}_sY_{\ell m}(\th,\vf)=&\left[(\ell -s)(\ell +s+1)\right]^{1/2}{}_{s+1}Y_{\ell m}(\th,\vf), \\
\bar\eth {}_sY_{\ell m}(\th,\vf)=&-\left[(\ell +s)(\ell -s+1)\right]^{1/2}{}_{s-1}Y_{\ell m}(\th,\vf).
\end{align}

Eq.\ \eqref{hertz} can be inverted to find the desired $\Psi$. In particular for circular orbits an algebraic mode by mode inversion is possible for Eqs.\ \eqref{hertz0} and \eqref{hertz4}:
\begin{subequations}
\label{invhertz}
\begin{align}
\Psi^{\rm ORG}_{\ell m}=&8\frac{(-1)^m(\ell +2)(\ell +1)\ell (\ell -1)\bar\psi_{0\,\ell ,\, -m}+12 iMm\omega\psi_{0\,\ell m}}{[(\ell +2)(\ell +1)\ell(\ell -1)]^2-144M^2 m^2\omega^2},\\
\Psi^{\rm IRG}_{\ell m}=&8\frac{(-1)^m(\ell +2)(\ell +1)\ell (\ell -1)\bar\psi_{-2\,\ell ,\, -m}-12 iMm\omega\psi_{-2\,\ell m}}{[(\ell +2)(\ell +1)\ell(\ell -1)]^2-144M^2 m^2\omega^2},
\end{align}
\end{subequations}
where $\psi_{-2}\equiv \rho^{-4}\psi_4$. We have denoted $\Psi_{\ell m}$ the modes of the radial part of the full Hertz potential and consistently for the scalars $\psi_0$ and $\psi_{-2}$.
\subsection{Non-radiative modes}\label{lowmodes}
The reconstruction from Weyl scalars recovers the full gauge invariant radiative part of the solution (namely the $\ell\geq 2$ sector). Wald showed that the solution needs to be completed by including corrections to the Kerr mass and angular momentum \citep{waldtheo}. Wald also allowed the inclusion of perturbations to other algebraically special solutions (C-metrics and Kerr-NUT metrics) and he proved that they are not physical in vacuum. Friedman {\it et al.} showed that the C and Kerr-NUT perturbations can be ruled out in the vacuum spacetime outside the trajectory of a point particle \citep{friedman1}.

The shift in the mass parameter across the $r=r_0$ surface is encoded in the monopole part of the solution (the $\ell =0$, $m=0$ mode). In the Lorenz gauge the nonvanishing components of this perturbations are \citep{Barack:2005nr}  
\begin{subequations}\label{l0m}
\begin{align}
h_{tt}^{\ell =0}(r\leq r_0)=&-\frac{A f M P(r)}{r^3},\\
h_{rr}^{\ell =0}(r\leq r_0)=&\frac{A M Q(r)}{r^3 f},\\
h_{\th\th}^{\ell =0}(r\leq r_0)=&\sin^{-2}\th h_{\vf\vf}^{\ell =0}(r\leq r_0)=A f M P(r),
\end{align}
\end{subequations}
where
\begin{align}
A=\frac{2\mathsf{m}{\cal E}}{3Mr_0 f_0}&\left[M-(r_0-3M)\ln f_0\right],\\
P(r)=r^2 +2Mr+4M^2,\qquad & Q(r)=r^3-Mr^2-2M^2r+12M^3,
\end{align}
and $f_0\equiv f(r_0)$. The external components are 
\begin{subequations}\label{l0p}
\begin{align}
h_{tt}^{\ell =0}(r\geq r_0)=&\frac{2\mathsf{m}{\cal E}}{3r^4r_0 f_0}\left\{\frac{}{} 3r^3(r_0-r)+M^2(r_0^2-12Mr_0+8M^2)+ \right. \nonumber \\
	&\left. (r_0-3M)\left[-rM(r+4M)+rP(r)f\ln f+8M^3\ln\left(\frac{r_0}{r}\right)\right]\right\},\\
h_{rr}^{\ell =0}(r\geq r_0)=&-\frac{2\mathsf{m}{\cal E}}{3Mr^4r_0 f_0f^2}\left\{\frac{}{}-r^3r_0-2Mr\left(r_0^2-6Mr_0-10M^2\right)+ \right. \nonumber \\
   &\left. 3M^2 \left(r_0^2-12Mr_0+8M^2\right)+(r_0-3M) \left[5Mr^2+\frac{r}{M}Q(r)f\ln f-8M^2(2r-3M)\ln\left(\frac{r_0}{r}\right)\right]\right\},\\
h_{\th\th}^{\ell =0}(r\geq r_0)=&\sin^{-2}\th h_{\vf\vf}^{\ell =0}(r\geq r_0)=-\frac{2\mathsf{m}{\cal E}}{9rr_0 f_0}\left\{\frac{}{}3r_0^2M-80M^2r_0+156M^3 \right. \nonumber\\
 	&\left. +(r_0-3M)\left[-3r^2-12Mr+3\frac{r}{M}P(r)f\ln f+44M^2+24M^2\ln\left(\frac{r_0}{r}\right)\right]\right\}.
\end{align}
\end{subequations}
Notice that as $r\to \infty$ the $tt$ component of the metric tends to a constant value, i.e., the metric is not asymptotically flat. Detweiler and Poisson showed \citep{DetPoi} that the Lorenz-gauge metric given by Eqs.\ \eqref{l0m} and \eqref{l0p} is unique and any gauge transformation within the class of Lorenz gauges would make the metric singular at infinity, at the horizon or in both limits. This pathology of the metric can be cured by moving away from the Lorenz gauge by performing a shift $t\to t(1+\alpha)$ with constant $\alpha\sim O(\mathsf{m})$. It is straight forward to show using Eq.\ (6) of \cite{barack1} that this gauge transformation does not contribute to the values of the SF.

For $\ell =1$, $m=0$ there is only one non-vanishing component of the MP \citep{Barack:2005nr} 
\begin{align}
\label{l1m0}
h_{t\varphi}^{\ell =1,m=0}(r)=-2\mathsf{m}{\cal L} \sin^2 \th\left[\frac{r^2}{r_0^3}\Theta(r_0-r)+\frac{1}{r}\Theta(r-r_0)\right],
\end{align}
where $\Theta$ is the usual step function.

We can calculate the contribution to the retarded force from the $\ell=0,1$ solutions by directly substituting \eqref{l0m}, \eqref{l0p} and \eqref{l1m0} in Eq.\ \eqref{force}. The resulting contribution to the force agrees with the values first obtained by Detweiler and Poisson \citep{DetPoi} at $\theta=\frac{\pi}{2}$.

The $\ell =1$, $m=1$ mode can be added numerically using the prescription described in \citep{DetPoi}. This mode is related to the motion around the center of mass of the BH-particle system. A detailed physical interpretation and comparison with a Post-Newtonian calculation can be found in \citep{DetPoi}.

\section{Numerical implementation for circular orbits}\label{num}
\subsection{Algorithm}\label{Algo}

The algorithm to obtain numerically the GSF in a Schwarzschild background follows the one used by Shah {\it et al}.\ \citep{friedman2}, except when stated. We outline the steps of our numerical implementation here.

\begin{itemize}
\item Choose the orbit at radius $r_0$. Obtain the relevant orbital parameters ${\cal E}$, ${\cal L}$ and $\Omega$ using Eq.\ \eqref{orbpar}. We fix the maximum number of modes to compute, $\ell_{\rm max}=80$. This choice of $\ell_{\rm max}$ guarantees convergence and provides enough $\ell$-modes to fit the $\ell>\ell_{\rm max}$ contribution (described in Sec.\ \ref{Renorm_scheme} bellow) without introducing numerical noise or becoming computationally expensive. 

\item For each static mode of the ORG with $\ell\geq 2$ %with $m=0$ 
we analytically calculate the radial function $R_{0}(r)$ via Eq.\ \eqref{PlQl} [for the IRG we calculate $R_{4}(r)\equiv r^4f^2 \bar{R}_{0}(r)$].

\item For each $m \neq 0$ we numerically integrate the radial Sasaki-Nakamura equation in $r_*$ with suitable boundary conditions (see appendix \ref{SN}). The integration routine returns the value of the function and the first derivative with respect to $r_*$. We algebraically relate the solutions $R_{4}(r)\equiv r^4f^2 \bar{R}_{0}(r)$ at the particle's location and calculate higher order derivatives using the radial part of Teukolsky equation. We also find the homogeneous solutions using the MST method described in the next section. The agreement between the two methods will be discussed in Sec.\ \ref{comp}.

\item We construct the inhomogeneous solutions using the standard variation of parameters method, imposing the jump conditions for the homogeneous solutions and their first derivatives at $r= r_0$, using the gravitational source. Shah {\it et al}.\ \citep{friedman2} performed an analytic integration of the Green's function over the source terms to construct the particular inhomogeneous solution $\psi_0$. We have checked that these two methods are equivalent and leave no ambiguity in the value of the Weyl scalars. The resulting field $\psi_0(r)$ (and $r^{4}\psi_4(r)$) is discontinuous at the location of the particle.

\item With the field $r^{4}\psi_4(r)$ [or $\psi_0(r)$ in the ORG] we find the harmonics of the Hertz potential $\Psi_{\ell m}^{\rm IRG}(r)$ [or $\Psi_{\ell m}^{\rm ORG}(r)$] using Eq.\ \eqref{invhertz}. The total Hertz potential can be computed as a sum over all modes with the corresponding angular and time dependence: $sY_{\ell m}(\theta,\varphi)\; e^{-i\omega t}$.

\item The MP can be recovered in the radiation gauge using Eq.\ \eqref{hIRG} [or Eq.\ \eqref{hORG}]. In particular we do the reconstruction for each $\ell$ and $m$.

\item  We calculate the $\ell$-modes of the full force $F^{\ell}_{full}$ (for each $\ell\geq 2$) by taking derivatives of the components of the $\ell$-modes of the Hertz potential [$\ell$-modes of Eq.\ \eqref{force} in Boyer--Lindquist coordinates acting on the $\ell$-modes of Eq.\ \eqref{hORG} for the ORG and Eq.\ \eqref{hIRG} for the IRG]. This is a convenient way for recording the contributions with respect of their angular dependence on ${}_sY_{\ell m}(\theta,\varphi)$ with $s=\pm 2, \pm 1, 0$ for the posterior re-expansion in terms of the usual scalar spherical harmonics. The explicit expressions are given in Eq.\ \eqref{Fr16ORG} and  Eq.\ \eqref{Fr16IRG}.

\item The remaining modes $\ell=0,1$ are added in the LG as discussed in Sec.\ \ref{lowmodes}. A method for including the low modes in the case of eccentric orbits in Kerr will be presented in a following paper \citep{BMP2}.
 
\item We use the definitions of spin-weighted spherical harmonics in terms of derivatives of scalar spherical harmonics [See Eq.\ \eqref{sYtoY} in the appendix]. This way we can implement the appropriate coupling formulas \citep{BaSa} to re-express the $r$ component of the retarded force in the basis of the scalar spherical harmonics where the mode-sum was derived \cite{baracksf, barackelgr}. In Schwarzschild the coupling is finite and it relates a given $\ell$-mode with its four nearest ``neighbours'', namely, contributions to a given $\ell$  spherical harmonic mode come from the $\ell \pm 2,\ell \pm 1$ and $\ell$ spin-weighted modes. The latter implies that we need to calculate $\ell_{\rm max} +2$ modes to have all the contributions to the $\ell_{\rm max}$ term in the mode-sum. This coupling and the implementation of the average mode-sum formula were missing in the prescription described in \citep{friedman2}.  

\item After all the contributions to a single $\ell$-mode are considered we apply the mode-sum regularization formula given by Eq.\ \eqref{modesumave} to obtain the radial component of the SF.  

\item We extrapolate the remaining $\ell > \ell_{\rm max}$ modes doing a numerical fitting of the regularized modes included in the mode-sum formula as described in Sec.\ \ref{Renorm_scheme}.
\end{itemize}

Once the MP are computed we relate the temporal component of the SF to the total flux of energy $\dot{{\cal E}}_{\rm rad}\equiv -{\mathsf{m}}\frac{d{\cal E}}{dt}=\frac{F_t}{u^t}$ according to \cite{detweiler, Barack:2009ux}
\begin{equation}\label{tcomp}
F^t= \sum_{\ell, m}\frac{im\Omega\mathsf{m}}{2 f}u^\alpha u^\beta h_{\alpha\beta}^{\ell m}.
\end{equation}
where $ h_{\alpha\beta}^ {\ell m}$ are the harmonic modes of the MP in the basis of spin-weighted spherical harmonics. The sum in Eq.\ \eqref{tcomp} converges exponentially fast and does not require regularization.

\subsection{MST (Mano-Suzuki-Takasugi) method}\label{MSTmet}

To calculate the solutions to the radial part of the homogeneous Teukolsky equation, we also use the \emph{MST-method} \cite{MST, SFW}. In this method, instead of numerically integrating the equation from the boundaries (infinity and event horizon), they are written as a sum over known analytic functions: the ingoing solution $R_H$ (which is regular at the event horizon) is written as a sum over hypergeometric function ($\,_2F_1$) and the outgoing solution $R_\infty$ (regular at infinity) is written as a sum over (Tricomi's) confluent hypergeometric function ($U$),

\begin{align} \label{MSTeqn}
R_H &= e^{i\epsilon x}(-x)^{-2-i\epsilon} \sum_{n=-\infty}^{n=\infty} a_n \,_2F_1\left(n+\nu+1-i\epsilon,-n-\nu-i\epsilon,-1-2i\epsilon;x\right), \nonumber \\
R_\infty &= e^{iz} z^{\nu-2} \sum_{n=-\infty}^{n=\infty} (-2z)^n b_n U(n+\nu+3-i\epsilon,2n+2\nu+2;-2iz),
\end{align}
where $x = 1-\frac{r}{2M}$, $\epsilon = 2Mm\Omega$ and $z=-\epsilon x$. We refer the readers to \cite{MST, SasTakLivRev} for the calculation of the parameter $\nu$ (\emph{renormalized angular momentum}), and the coefficients $a_n$ and $b_n$. The solutions were calculated with 16-35 digits of accuracy\footnote{The solutions are less accurate near the event horizon but achieve high accuracy as we move further away.} for orbital radii ranging from $r_0=6M-200M$, respectively.% 

\subsection{Fitting the large-$\ell$ tail}\label{Renorm_scheme}

In the discontinuous radiation gauge where the radiative modes of the SF are calculated from the Weyl scalar in the limit $r \rightarrow r_0^{\pm}$, we find that the singular part of the SF contains odd, negative powers of $L=(\ell+1/2)$ on either side\footnote{If the tail was fitted using the averaged modes of the retarded force only even powers of $L$ would appear.} of $r_0$. Each of the side dependent values required in the averaged version of the mode-sum are computed according to 
\begin{align}\label{tail}
F^\alpha_{\pm} = \sum_{\ell=0}^{\ell_\textrm{max}} \left[\left(F^{\alpha}_{\rm full}\right)_\pm^\ell \mp A^\alpha L - B^\alpha \right] - D^\alpha_\pm+ \sum_{\ell_\textrm{max}+1}^\infty \left[ \frac{\tilde{E}_2^\pm}{L^2} + \frac{\tilde{E}_4^\pm}{L^4} + \frac{\tilde{E}_5^\pm}{L^5} + \frac{\tilde{E}_6^\pm}{L^6} + \cdots + \frac{\tilde{E}_{k_\textrm{max}}^\pm}{L^{k_\textrm{max}}} \right] + O\left(\frac{1}{ \ell_\textrm{max}^{k_\textrm{max}}}\right),
\end{align}
where the $\pm$ superscript indicates that the fitting parameters are calculated using the side dependent values of $\left[\left(F^{\alpha}_{\rm full}\right)_\pm^\ell \mp A^\alpha L - B^\alpha \right]$ and in general $\tilde E_{k}^{+} \neq \tilde E_{k}^{-}$. We extract the coefficients $\tilde{E}_k^\pm$ by matching $\left[ \left(F^\alpha_{\rm full}\right)_\pm^\ell - A_\pm^\alpha L - B^\alpha \right]$ (from a certain $\ell_\textrm{min}$ to $\ell_\textrm{max}$) to a power series of the form\footnote{In the Lorenz gauge a series of the form $E_2/((2\ell -1)(2\ell +3))+E_4/((2\ell - 3)(2\ell -1)(2\ell +3)(2\ell +5))...$ is used to fit the singular part of the force and increase the convergence rate \cite{DetMesWhi}. Analytical expression for $E_2$, $E_4$, $E_6$ were given in \cite{Heff} and we verify that they have different values than the parameters we would obtain by fitting the averaged modes to a similar series.}
\begin{align}\label{tail2}
\frac{\tilde{E}_2^\pm}{L^2} + \frac{\tilde{E}_4^\pm}{L^4} + \frac{\tilde{E}_5^\pm}{L^5} + \frac{\tilde{E}_6^\pm}{L^6} + \cdots + \frac{\tilde{E}^\pm_{k_\textrm{max}}}{L^{k_\textrm{max}}}.
\end{align}

The best-fit values of $\tilde{E}^\pm_k$ are extracted by modifying $\ell_\textrm{min}$ and $k_\textrm{max}$ using the procedure described in \cite{friedman2}. The SF is then calculated using Eq.\ \eqref{modesumave}, where the $\ell > \ell_{\rm max}$ tail is included using the best numerical fit.

An interesting detail to be noted here is that we numerically find $F^\alpha_{\rm self}$ to be independent its mode decomposition --- whether written as a sum over mixed spin-weighted spherical harmonics as done in Eqs.\  \eqref{Fr16IRG} and \eqref{Fr16ORG} or as a sum over ordinary spherical harmonics as done in Eq.\ \eqref{ForceYlm}--- unlike the sided limits $F^\alpha_\pm$. We will further comment on this numerical result on Sec.\ \ref{comp}. 

\section{Results}\label{results}
\subsection{Convergence of the mode sums for $F^r$ and $F^t$}

A feature of the mode-sum regularization procedure is that it provides an immediate validity test of the results. If the retarded values of the force and the implementation of the coupling formulas that allow us to express the force as purely spherical harmonics contain a systematic error, then the sum over $\ell$-modes after regularization may not converge to the physical value of the SF\footnote{Convergence might still occur, for example \cite{friedman2} where the re expansion to scalar harmonics and the average were not included.}. It is also required to consistently use the extension of the four-velocity [we used the same extension as the LG regularization parameters $A^\alpha$ and $B^{\alpha}$ of Eq.\ \eqref{regpar}]---in principle we can also extend the components of the metric and connection terms--- when calculating the retarded values and the regularization parameters, otherwise the mode-sum will not give the correct value. 
\begin{figure}[htp]
  \centering
 
\includegraphics[width=85.5 mm]{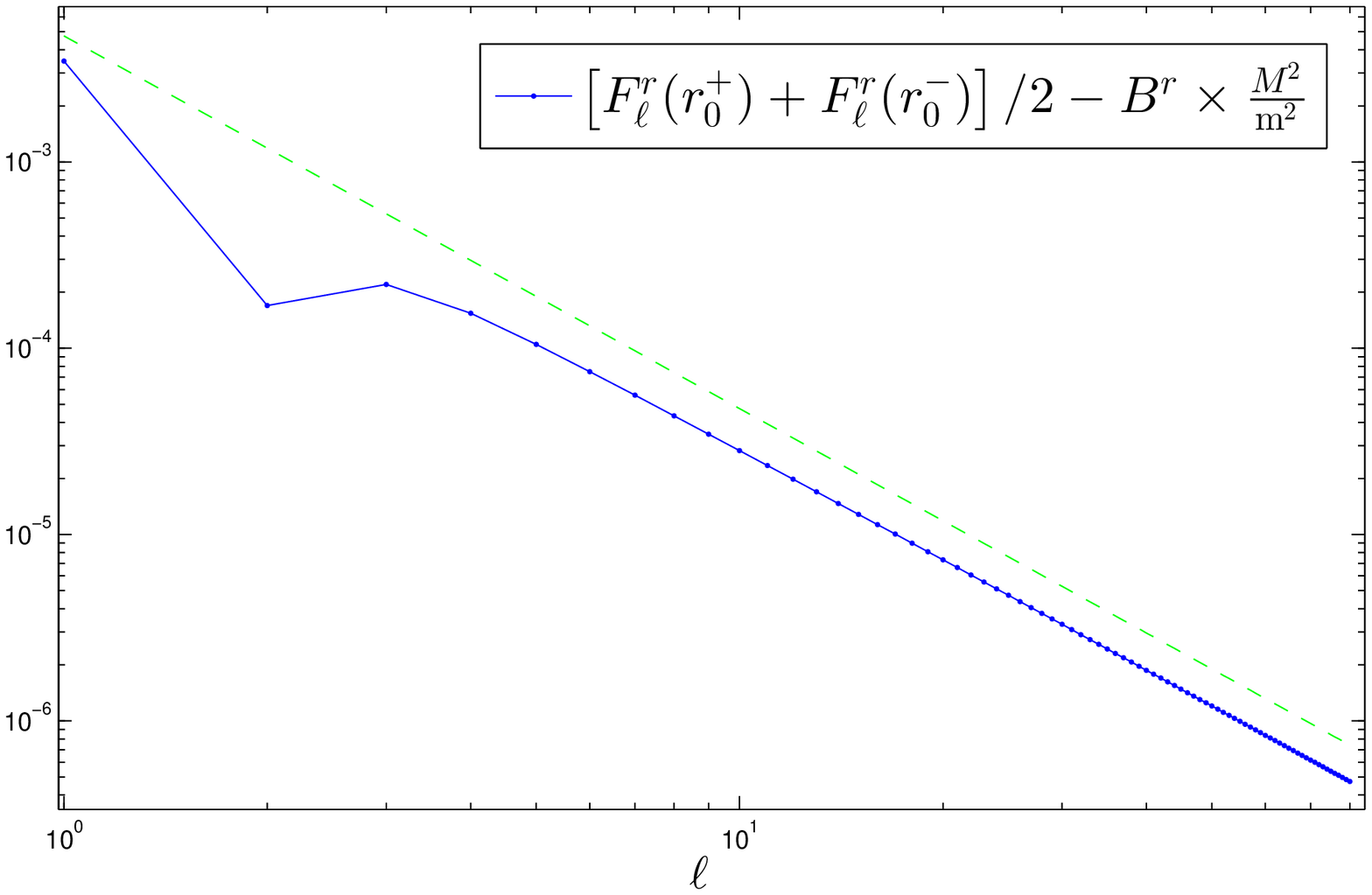} 
\includegraphics[width=85.5 mm]{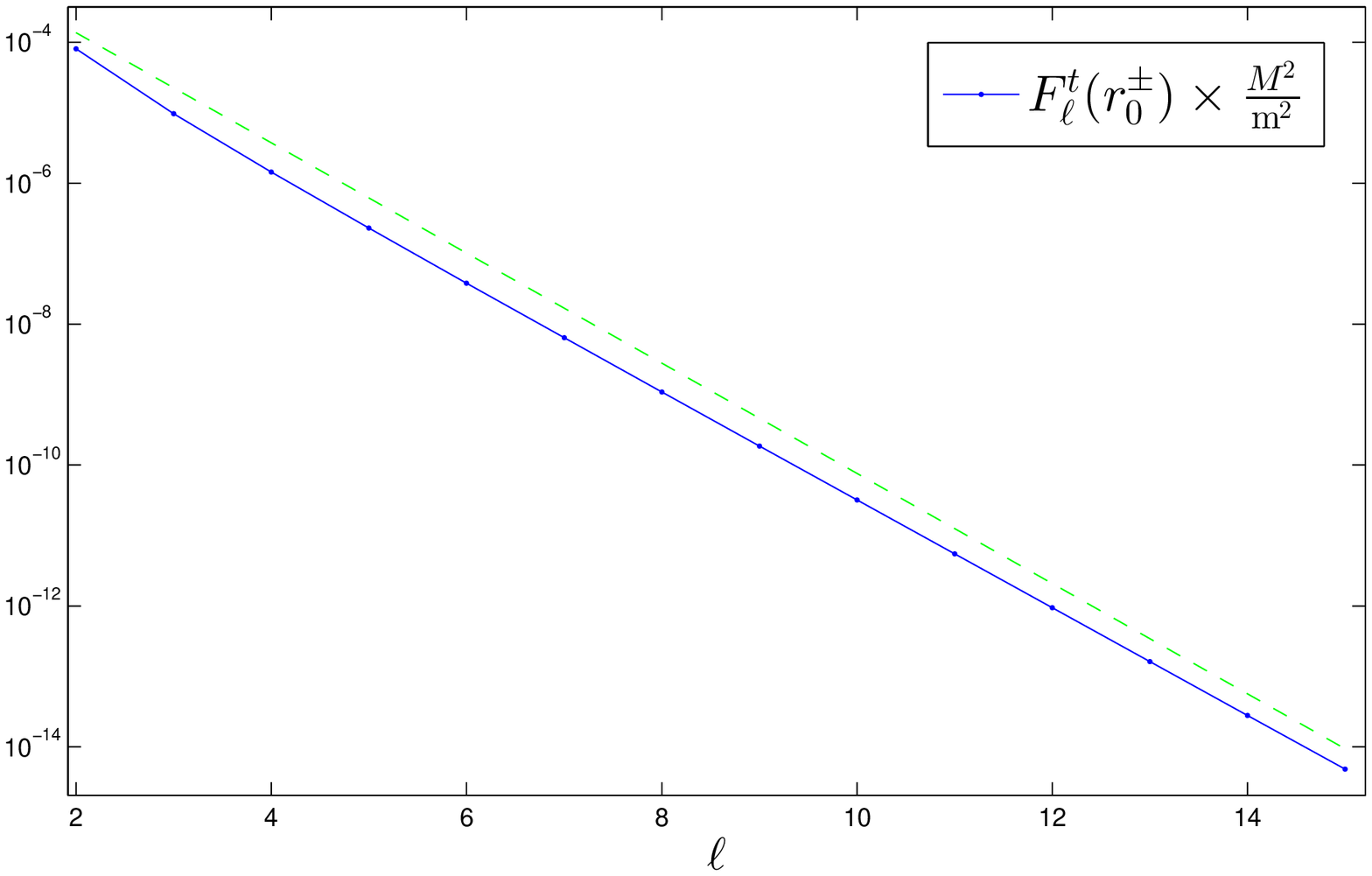}
 
  \caption{Left Panel shows the convergence of the mode sum for the {\it r} component of the SF (solid blue line in log-log scale) computed using the average version of the mode-sum formula [Eq.\ \eqref{modesumave} with $\ell_{\rm max} =80$, only $A^r L$ and $B^r$ are subtracted]. The reference line (green dashed) corresponds to the $1/\ell^2$ fall off at large $\ell$. The right panel shows the convergence of the {\it t} component (solid blue line in semi-log scale) of the SF, we show only $\ell=15$ modes. In this case the reference line (green dashed) shows exponential convergence. In both cases the results correspond to an orbital radius of $r_0=10M$.}\label{conv}

\end{figure}

For the radial component (left panel of Fig.\ \ref{conv}) we found that the sum over $\ell$ modes of the average $\frac{1}{2}\left[F^r_\ell(r_0^+)+F^r_\ell(r_0^-)\right]$ converges $\sim 1/\ell$, with the green (dashed) line as reference. In the case of the time component (right panel of Fig.\ \ref{conv}), we show the exponential convergence of the sum.% manifests already at $\ell \sim 15$. 

\subsection{Flux of energy}
We calculate the fluxes at infinity ($-\mathsf{m}{\cal E}_{ \infty}$) and at the event horizon ($-\mathsf{m}{\cal E}_{\rm EH}$) following the procedure given in \citep{hughes}. And we verify numerically that
\begin{align}\label{balance}
\frac{d{\cal E}}{dt}=\frac{d{\cal E}_{\rm EH}}{dt}+\frac{d{\cal E}_{ \infty}}{dt},
\end{align}
is satisfied up to $\sim 10^{-5}$ of relative difference for all radii.

Our results are consistent with previous works by Barack and Sago \citep{BaSa}, and more recently Gundlach {\it et al}.\ \citep{sarpgundlach}. Our calculation shows that at the Innermost Stable Circular Orbit (ISCO) the ratio $\dot {\cal E}_{\rm EH}/\dot {\cal E}_{\infty}$ has a value of $3.27\times 10^{-3}$ and decreases monotonically with $r_0$ up to $2.06\times 10^{-9}$ when $r_0=150M$.

\subsection{Comparison of results}\label{comp}
\begin{figure}
  \centering
  
\includegraphics[width=17 cm]{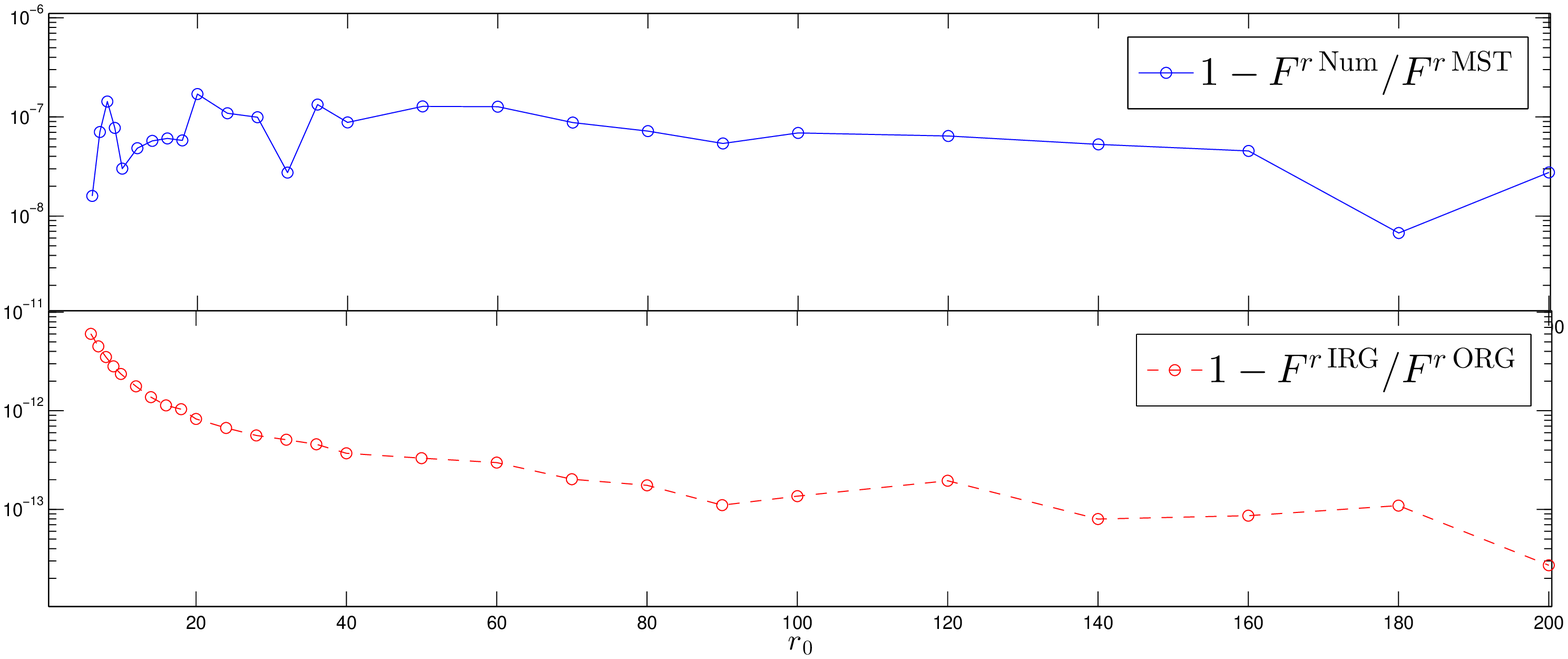} 

  \caption{Relative difference for the averaged $r$ component of the SF. The blue (solid) line compares the values in the ORG computed through numerical integration of the Sasaki-Nakamura field against the values calculated using the MST analytical method. The estimated error of the numerical method is dominated by the $\ell >\ell_{\rm max}$ fitted term, while the error of the MST method is given by the inclusion of the even dipole mode, these errors are shown explicitly in Table \ref{rvalues} at the end of Appendix \ref{IRGvsORG}. The red (dashed) line compares the relative difference up to the required accuracy between the force calculated from the IRG and the ORG modes.}\label{SvsM}
\end{figure}

We now present a comparison between the radial component of the SF calculated using the MST method and the numerical integration of Sasaki-Nakamura function. Fig.\ \ref{SvsM} shows in blue (solid) line the fractional error in $F^r(r_0)$ for a sample of radii, taking the values calculated with the MST method as more accurate. Such values are obtained using Eq.\ \eqref{modesumave} with $80$ calculated modes (as described in Sec.\ \ref{Algo}) and a fitted tail of the form given by Eq.\ \eqref{tail2} on each sided limit. In red (dashed line) we show the fractional difference between the IRG and the ORG values. In this case both results were obtained by using the Sasaki-Nakamura method. The values used to generate the plot can be found in Table \ref{rvalues} in Appendix \ref{IRGvsORG}.

A similar table was presented in the mentioned work by Shah {\it et al}.\ \citep{friedman2}, but the values for the SF don't agree with ours. The computation in \citep{friedman2} differs from the one we have presented here in several ways as we have briefly stated in previous sections. We now summarize the differences and discuss why our values correspond to the physical SF calculated using the RG modes. The table in \citep{friedman2} has the values of the sum after regularisation of the $\ell$-modes in the basis of spin-weighted spherical harmonics. Even though regularization is possible at the level of `any' harmonic basis, as we mentioned before, the LG regularization parameters used in \citep{friedman2} are only suitable for modes expressed in terms of the usual spherical harmonics, just like we have done in the present work. A numerical experiment showed that using the averaged version of the mode-sum method gives the same results in either basis (spin-weighted and scalar spherical harmonics) in the case of circular orbits around Schwarzschild, but this might not happen in more general cases. Only re-expanding the $\ell$-modes in terms of the usual spherical harmonics or deriving regularization parameters for a different basis will give the correct SF. A second difference relies on the average version of the mode-sum formula introduced in \citep{BMP1}: the values of \citep{friedman2} correspond only to the one sided upper limit (when $r\rightarrow +r_0$), in which case the inclusion of a non-trivial $D^\alpha$ parameter in the appropriate extension is required\footnote{The rigid extension was used in \citep{BMP1}.} to give the SF (a different numerical value to the averaged) in a half string RG or a half-string locally Lorenz gauge, where the motion is not well defined \citep{BMP1}. We are aware of an {\it erratum} soon to be presented from the authors of \citep{friedman2} clarifying the issues we have raised in this paper.

In principle the SF in the ORG and the IRG could have different values. In fact by just looking at Eqs.\ \eqref{Fr16IRG} and \eqref{Fr16ORG} it is not obvious that the results would agree. The Hertz potential $\Psi$ takes a different form when calculated in the ORG and IRG. For circular equatorial orbits around a Schwarzschild BH it turns out that the MP and the values of the SF in the IRG and ORG give the same value. The equivalence of the MP in both gauges can be shown analytically using the symmetries of Teukolsky equation. This agreement has also been confirmed numerically up to the required accuracy.

A LG code for circular orbits of Schwarzschild calculates the SF in the strong field regime in approximately 2 hours with $\ell_{\rm max}\sim 25$ and a factional accuracy of $\lesssim 10^{-4}$ \cite{BaSa}. Our numerical integration can achieve the same accuracy ($\lesssim 10^{-4}$) running on a single core in about 45 minutes and an accuracy of $\lesssim 10^{-12}$ in about 1.5 hours calculating $\ell_{\rm max}\sim 25$ modes. With the MST method we calculate typically 85 modes with an accuracy of 16-35 digits (Sec.\ \ref{MSTmet}) within 6-19 hours (we require more time in the strong field regime) running in 16 processors.

The values of the radial component of the SF in the LG \cite{BaSa} asymptotically agree with the values given in Table I. This in not surprising since the change in the force due to the gauge transformation from Lorenz to ORG falls off at least $\sim r^{-3}$ (see Eq.\ (A25) of \citep{friedman1}). %A direct comparison between the SF values is a non trivial task. One might be tempted to calculate $\delta D^{\pm}_\alpha$ for the appropriate extension is the rigid one.} on each of the sided limits, this will give the force in the half string gauges which are regular inside and outside the orbit. In this half string gauges .

\subsection{Sources of numerical error}
The total value of the radial component of the SF has two pieces. The first one $F_{r}^{\ell \leq \ell_{\rm max}}$ is obtained by the methods already described. The remaining {\it tail} piece $F_{r}^{\ell > \ell_{\rm max}}$ is extrapolated numerically as described in Sec.\ \ref{Renorm_scheme} using $\tilde N=\ell_{\rm max}-\ell_{\rm min}$ of the regularized large $\ell$-modes. We checked that our solutions are insensitive to variations in the numerical parameters to the required accuracy.

The error in calculating the radial parts of the homogeneous Teukolsky equation using the MST-method can be reduced by first, numerically calculating $\nu$ with a very high accuracy (usually higher than the one mentioned in Table I), and second, by choosing a high enough $n_\textrm{max}$, the cut-off in $n$-series of the hypergeometric and confluent hypergeometric series in Eq.\ \eqref{MSTeqn}. To reduce the computation-time, we find relations between the derivatives of the hypergeometric and confluent hypergeometric functions appearing in Eq.\ \eqref{MSTeqn} using a combination of various Guass's relations for contiguous functions. 

The numerical integration of Sasaki-Nakamura equation is done using a modification to complex variables with quadrupole precision of the adaptive stepsize Bulirsch-Stoer routine described in \cite{NumRec}. We allow a relative error of $1/10^{15}$ on each step of the integration. These errors propagate to give a relative error $\sim 1/10^{12}$ in the value of each harmonic of the Sasaki-Nakamura field and its first derivative. However these systematic errors are subdominant with respect of the contributions from the {\it tail}.

The accuracy to which the coefficients $\tilde{E}_k^\pm$ in Eq.\ \eqref{tail} can be extracted depends on $\tilde N$ and the accuracy of the regularized modes. Due to its high accuracy the MST method allows a very accurate extrapolation of the tail. With respect of the values reported in Table \ref{rvalues} the total tail accounts for the last 4-5 digits of agreement between the Sasaki-Nakamura and MST methods. The relative difference of the two methods is within the error bars reported for the computation made using numerical integration. These error bars were estimated by varying the numerical parameters of the fitting. The error bars for the MST method values were estimated from varying the inner boundary of the integration [$r_{\rm min}=(2+\epsilon) M$] of the dipole even mode from $\epsilon= 10^{-9}$ to $\epsilon= 10^{-6}$, which dominates over the accuracy of the MST modes (16-35 significant digits as mentioned in Sec.\ \ref{MSTmet}). 

\section{Concluding remarks}
In this work we have presented for the first time a full calculation of the gravitational SF from the radiation gauge MP. We have also shown the equivalence (at the level of SF calculation for circular orbits around Schwarzschild) of working in an IRG or an ORG, made a successful comparison between the MST method and numerical integration of Teukolsky equation, and have tested the numerical code by calculating well know quantities available in the literature, such as the energy fluxes and the {\it t} component of the SF.

An extension of this computation using the MST method will soon follow for general orbits around a Kerr background. Teukolsky equation remains separable in Kerr ---unlike the tensorial equations in the LG--- and the metric reconstruction procedure is well understood. One of the challenges in SF calculations of more general orbits (both in Schwarzschild and Kerr) is the inclusion of the mass and angular momentum perturbations that complete the reconstructed MP \citep{BMP2}. A second challenge in the Kerr calculations is the re-expansion of the $\ell$-modes into the spin-0 spherical harmonics. This involves a numerical projection of the spin-weighted spheroidal harmonics (in which the harmonics modes of the full force is obtained) which might not have a finite coupling as they exhibited in the present work. The coupling will be simpler if a suitable off the worldline extension of the four velocity is chosen.

\section{Acknowledgements}
The authors would like to thank Leor Barack for many productive discussions and a careful reading of this paper. We thank Adam Pound, John Friedman, Sarp Akcay, Norichika Sago and Scott Hughes for useful comments. We also thank an anonymous referee for useful suggestions that we included in the final version. CM acknowledges support from CONACyT. AS work is funded by the European Research Council under the European Union’s Seventh Framework Programme FP7/2007-2013/ERC Grant No.\ 304978.
%%%%%%%%%%%%%%%%%%%%%%%%%%%%%%%%%%%%%%%%%%%%%%%%%%%%%%%
\appendix

\section{Static modes}\label{asolm=0}
For the static modes ($m=\omega=0$) we have two linearly independent solutions proportional to associated Legendre polynomials of first and second kind:
\begin{subequations}
\label{PlQl}
\begin{align}
R_{0-}(r)\equiv &\frac{{\sf P}_{\ell}^2\left(\frac{r-M}{M}\right)}{r(r-2M)}=-\frac{\Gamma(\ell +3)}{16M^2 r\Gamma(\ell -1)}{}_2F_{1}\left[2-\ell,\ell +3;3,-\frac{r}{2M}\right],\\
R_{0+}(r)\equiv &\frac{{\sf Q}_{\ell}^2\left(\frac{r-M}{M}\right)}{r(r-2M)}= \frac{2^\ell\Gamma(\ell +3)\Gamma(\ell +1)}{M^2 r\Gamma(2\ell +2)}\left(\frac{r-M}{M}\right)^{-\ell -3}{}_2F_1\left[\frac{\ell}{2}+2,\frac{\ell +3}{2};\ell +\frac{3}{2},\frac{M^2}{(r-M)^2}\right],
\end{align}
\end{subequations}
where ${}_2F_1$ are the hypergeometric functions.
$R_{0-}(r)$ is regular at the horizon but diverges as $r^{\ell-2}$ at infinity for any $\ell >2$. When $r\to \infty$ and $\ell =2$, $R_{0-}$ goes to a constant. $R_{0+}(r)$ is not regular at the horizon since it behaves $\sim (r-2M)^{-2}$, but is regular at infinity where its leading order is given by $r^{-\ell-3}$. The asymptotic behaviour of these solutions was previously discussed by Barack and Ori \citep{leoranasol} near the event horizon and by Poisson \citep{poissonansol} and Keild {\it et al}.\ \citep{Keidlhom}.

\section{Chandrasekhar--Sasaki-Nakamura transformation}\label{SN}
In the Schwarzschild case the radial part of Sasaki-Nakamura equation reduces to 
\begin{align}
\label{RW}
\left[\frac{d^2}{dr_*^2}+\omega^2 -V(r)\right]X_{\ell m}(r)=0, \quad {\rm with} \quad V(r)\equiv f\left(\frac{r\ell(\ell+1)-6M}{r^3}\right).
\end{align}
The relation between the solutions of the homogeneous Teukolsky equation with $s=-2$ and the function $X(r)$ was first found in \citep{sasnak}. In Schwarzschild it can be written as 
\begin{align}
R_{\,4\ell m}(r)=2rf(r-3M+ir^2\omega)\frac{X'_{\ell m}(r)}{\eta}+\left[rf\ell(\ell+1)-6Mf-2r\omega(3iM-ir+r^2\omega)\right]\frac{X_{\ell m}(r)}{\eta},
\end{align}
where $\eta=(\ell -1)\ell(\ell+1)(\ell+2)-12iM\omega$ and the prime denotes derivatives with respect of $r$.
To integrate Eq.\ \eqref{RW} we set boundary conditions which are regular at infinity and at the event horizon \citep{friedman2}:
\begin{subequations}
\label{BCsn}\begin{align}
X^{H}=& e^{i\omega r_*}\sum_{n=0}^{n_{\rm max}}c_n\left(\frac{r}{M}-2\right)^n, \\
X^{\infty}=&e^{-i\omega r_*}\sum_{n=0}^{n_{\rm max}}d_n\left(\frac{M}{r}\right)^n,
\end{align}
\end{subequations}
with $c_n=d_n=0$ for $n<0$. The values of the coefficients $c_n$ and $d_n$ are calculated according to the recurrence relations
\begin{align}
c_n=&-\frac{i(n-3)M \omega}{2n(n+4iM\omega)}c_{n-3}+\frac{\ell(\ell +1)-(n-2)(n-3+12 iM\omega)}{4n(n+4iM\omega)}c_{n-2} \nonumber \\
&+\frac{\ell(\ell+1)-2n^2+5n-6-12i(n-1)M\omega}{2n(n+4iM\omega)}c_{n-1}\\
\label{dn}
d_n=&\frac{-i}{2nM\omega}\left[(n-3)(n+1)d_{n-2}+(\ell +n)(\ell -n+1)d_{n-1}\right].
\end{align}
and $n_{\rm max}$ is chosen so that the relative difference between the $n+1$ and the accumulated sum is smaller than $10^{-15}$.

\section{Explicit expressions for the source and the force using IRG and ORG modes}\label{IRGvsORG}
The source and self-acceleration in the ORG were previously presented in \cite{friedman1, friedman2}. We include the ORG expressions for completeness. We have identified and corrected small typos in the sources ---an independent check lead us to notice an incompatibility between the corresponding equations for the source in \citep{friedman1} and \citep{friedman2}. The authors of \citep{friedman3} choose $\theta=\pi /2$ in their expressions for the self-acceleration which makes difficult to read the full angular dependence required to change the basis from spin-weighted spherical harmonics to the usual spherical harmonics in which the mode-sum scheme guarantees to give the right value of the SF. We write the source of Teukolsky equation as a sum of three terms $T_{\pm 2}=T^{(0)}+T^{(1)}+T^{(2)}$ according the angular dependence on the particle's location of each term.

The explicit form --- in the Schwarzschild case--- of the source terms in the IRG is
\begin{subequations}
\label{sourceIRG}
\begin{align}
T^{(0)}=&-\sum_{\ell m}\frac{\mathsf{m} u^t f_0^2}{4}\delta(r-r_0)\left[(\ell -1)\ell (\ell +1)(\ell +2)\right]^{1/2}{}_{-2}Y_{\ell m}(\theta,\varphi)\,\bar Y_{\ell m}\left(\frac{\pi}{2},\Omega t_0\right),\\
T^{(1)}=&\sum_{\ell m}\frac{\mathsf{m}\Omega u^t f_0 r_0^2}{2}\left[if_0\delta'(r-r_0)-\left(m\Omega +\frac{4iM}{r_0^2}\right)\delta(r-r_0)\right]\left[(\ell -1)(\ell +2)\right]^{1/2} \nonumber \\
&{}_{-2}Y_{\ell m}(\theta,\varphi){}_{-1}\bar Y_{\ell m}\left(\frac{\pi}{2},\Omega t_0\right),\\
T^{(2)}=&\sum_{\ell m}\frac{\mathsf{m}\Omega^2 u^t r_0^4}{4}\left[f_0^2\delta''(r-r_0)+\left(2i m\Omega f_0-\frac{2 (r_0+2M)f_0}{r_0^2}\right)\delta'(r-r_0) \right.\nonumber  \\
&\left. -\left(m^2\Omega^2 +\frac{2im \Omega (r_0+M)}{r_0^2}-\frac{2(4M-r_0)}{r_0^3}\right)\delta(r-r_0)\right]{}_{-2}Y_{\ell m}(\theta,\varphi){}_{-2}\bar Y_{\ell m}\left(\frac{\pi}{2},\Omega t_0\right).
\end{align}
\end{subequations}
The corresponding source of the ORG is
\begin{subequations}\label{sourceORG}
\begin{align}
T^{(0)}=&-\sum_{\ell m}\frac{\mathsf{m} u^t}{r_0^4}\delta(r-r_0)\left[(\ell -1)\ell (\ell +1)(\ell +2)\right]^{1/2}\;{}_{2}Y_{\ell m}(\theta,\varphi)\bar Y_{\ell m}\left(\frac{\pi}{2},\Omega t_0\right),\\
T^{(1)}=&\sum_{\ell m}2\frac{\mathsf{m}\Omega u^t}{r_0^2}\left[i\delta'(r-r_0)+\left(\frac{m\Omega}{f_0}+\frac{4i}{r_0}\right)\delta(r-r_0)\right]\left[(\ell -1)(\ell +2)\right]^{1/2}{}_{2}Y_{\ell m}(\theta,\varphi){}_1\bar Y_{\ell m}\left(\frac{\pi}{2},\Omega t_0\right),\\
T^{(2)}=&\sum_{\ell m}\mathsf{m}\Omega^2 u^t\left[\delta''(r-r_0)+\left(\frac{6}{r_0}-\frac{2im\Omega}{f_0}\right)\delta'(r-r_0)\right. \nonumber \\
&\left.-\left(\frac{m^2\Omega^2}{f_0^2}+\frac{2im\Omega(3r_0-5M)}{r_0^2f_0^2}-\frac{10}{r_0^2}\right)\delta(r-r_0)\right]{}_{2}Y_{\ell m}(\theta,\varphi){}_2\bar Y_{\ell m}\left(\frac{\pi}{2},\Omega t_0\right).
\end{align}
\end{subequations}

The radial component of the full force in an IRG in terms of the tetrad component of the metric perturbation is given by
\begin{align} \label{ftetfull}
F^{r\,{\rm IRG}}_{\ell m} =  &(u^{t})^2 f\mathsf{m}\left[ \left( \frac{3}{4}{\bf D} + \frac{1}{2} {\bf \Delta} - \frac{M}{r^2 f}\right) h_{{\bf 22}} - \frac{M }{2\sqrt{2}r^2 f}\sin ^2\theta \left( \bar{\eth}_1h_{\bf 23} + \eth_{-1} h_{\bf 24} \right)+ \frac{i \Omega}{2f}\sin\theta \left( \eth_0 - \bar{\eth}_0 \right)h_{\bf 22} \right. \nonumber \\  
   &\left . - \frac{i\Omega r}{2\sqrt{2}} \sin\theta\left( {\bf \Delta} + \frac{2}{r} \right) (h_{\bf 23} - h_{\bf 24}) + \frac{M}{2\sqrt{2}r^2 f} \sin^2\theta\left( \eth_1 h_{\bf 23} + \bar{\eth}_{-1} h_{\bf 24}\right)  \right. \nonumber \\ 
  &\left. - \frac{M}{r} \sin^2\theta\left( \frac18 {\bf D} - \frac{1}{4f}{\bf \Delta} + \frac{1}{2r} \right) \left( h_{\bf 33} + h_{\bf 44}\right)- \frac{r \Omega^2}{f}\sin^2\theta h_{\bf 22}\right],
\end{align}

and the corresponding equation for the ORG:
\begin{align} \label{C6}
F^{r\,{\rm ORG}}_{\ell m}=  &-\frac{(u^{t})^2 f\mathsf{m}}{ r }\left[rf\left( \frac{f}{16}{\bf D} + \frac{3}{8} {\bf \Delta} - \frac{M}{2r^2}\right) h_{{\bf 11}} - \frac{M }{4\sqrt{2}r}\sin ^2\theta \left( \bar{\eth}_1h_{\bf 13} + \eth_{-1} h_{\bf 14} \right) + \frac{i rf\Omega}{8}\sin\theta \left( \eth_0 - \bar{\eth}_0 \right)h_{\bf 11} \right. \nonumber \\  
   &\left . - \frac{i\Omega r}{\sqrt{2}} \sin\theta\left( r{\bf \Delta} - \frac12 \right) (h_{\bf 13} - h_{\bf 14}) + \frac{M}{4\sqrt{2}r} \sin^2\theta\left( \eth_1 h_{\bf 13} + \bar{\eth}_{-1} h_{\bf 14}\right)  \right. \nonumber \\ 
  &\left. + M \sin^2\theta\left( \frac18 {\bf D} - \frac{1}{4f}{\bf \Delta} + \frac{1}{2r} \right) \left( h_{\bf 33} + h_{\bf 44}\right)- \frac{r^2f \Omega^2}{4}\cos^2\theta h_{\bf 11}\right].
\end{align}

The above equation differs from Eq.\ (44) of \cite{friedman2}, where the expression was calculated at $\theta=\pi/2$ and metric signature $(+,-,-,-)$. Eqs.\ \eqref{ftetfull} and \eqref{C6} have the exact powers of $\sin\theta$ to make it possible to write the final self-force as a finite sum over spin-0 ordinary spherical harmonics.

The radial component of the full force in the IRG can be computed as a sum of six terms for each value of $\ell$ and $m$ with different angular dependence:
\begin{subequations}
\label{Fr16IRG}
\begin{align}
F^r_{1\;\ell m}=&\frac{1}{4r_0^2}(u^t)^2\mathsf{m}\sqrt{(\ell-1)\ell(\ell+1)(\ell+2)} \left[f_0\partial_r +2\partial_t -\frac{2}{r_0}\left(f_0-\frac{M}{r_0}\right)\right](\Psi_{\ell m}+\bar{\Psi}_{\ell m})Y_{\ell m}(\theta,\varphi),\\
F^r_{2\;\ell m}=&\frac{1}{4 f_0r_0^4}(u^t)^2M\mathsf{m}\sqrt{(\ell-1)\ell(\ell+1)(\ell+2)} \left(r_0f_0\partial_r+r_0\partial_t -4f_0\right)(\Psi_{\ell m}+\bar{\Psi}_{\ell m})\sin^2\th \, Y_{\ell m}(\theta,\varphi),\\
F^r_{3\;\ell m}=&\frac{1}{4r_0^2}(u^t)^2\Omega i\mathsf{m}\sqrt{(\ell-1)(\ell+2)}\ell(\ell+1)(\Psi_{\ell m}-\bar{\Psi}_{\ell m}) \sin\th\left[\,{}_1Y_{\ell m}(\theta,\varphi)+\, _{-1}Y_{\ell m}(\theta,\varphi)\right],\\
F^r_{4\;\ell m}=&-\frac{1}{2f_0}(u^t)^2\mathsf{m} \Omega i\sqrt{(\ell -1)(\ell +2)}\left[\partial^2_t +2f_0\partial_t\partial_r +f_0^2\partial^2_r -\frac{2}{r_0^2}(M +r_0f_0)\partial_t -\frac{2f_0^2}{r_0}\partial_r\right.  \nonumber \\
 &\left.  +\frac{2f_0^2}{r_0^2}\right] (\Psi_{\ell m}-\bar{\Psi}_{\ell m})
 \sin\th\,{}_{-1}Y_{\ell m}(\theta,\varphi),\\
F^r_{5\;\ell m}=&\frac{1}{4 f_0r_0^4}(u^t)^2M\mathsf{m}(\ell -1)(\ell +2) \left(r_0f_0\partial_r+r_0\partial_t -2f_0\right)(\Psi_{\ell m}+\bar{\Psi}_{\ell m}) \sin^2\th\,_{-2}Y_{\ell m}(\theta,\varphi),\\
F^r_{6\;\ell m}=&-\frac{1}{4f_0^2r_0^5}(u^t)^2 M\mathsf{m}\left[r_0^4f_0\partial^2_t\partial_r +2r_0^4f_0^2\partial_t\partial^2_r +r_0^4f_0^3\partial^3_r +2r_0^3f_0^2\partial_t^2  +2r_0^2 f_0(r_0-5M)\partial_t\partial_r \right. \nonumber \\
 &\left. -2(r_0^2-6Mr_0+4M^2)\partial_t -2 r_0^2f_0^3\partial_r  \right](\Psi_{\ell m}+\bar{\Psi}_{\ell m}) \sin^2\th \,_{-2}Y_{\ell m}(\theta,\varphi),
\end{align}
\end{subequations}
where we have omitted to specify that $\Psi$ is the IRG hertz potential. The corresponding terms for the ORG are
\begin{subequations}
\label{Fr16ORG}
\begin{align}
F^r_{1\;\ell m}=&-\frac{1}{16}r_0f_0^2 (u^t)^2\mathsf{m}\sqrt{(\ell-1)\ell(\ell+1)(\ell+2)} \left[r_0f_0\partial_r-2r_0\partial_t+2\left(f_0+\frac{3M}{r_0}\right)\right](\Psi_{\ell m}+\bar{\Psi}_{\ell m})Y_{\ell m}(\theta,\varphi),\\
F^r_{2\;\ell m}=&-\frac{1}{16}f_0(u^t)^2M\mathsf{m}\sqrt{(\ell-1)\ell(\ell+1)(\ell+2)} \left[r_0f_0\partial_r-r_0\partial_t+2\left(1+f_0\right)\right](\Psi_{\ell m}+\bar{\Psi}_{\ell m})\sin^2 \th Y_{\ell m}(\theta,\varphi),\\
F^r_{3\;\ell m}=&\frac{1}{16}r_0^2f_0^2 (u^t)^2\Omega i\mathsf{m}\sqrt{(\ell-1)(\ell+2)}\ell(\ell+1)(\Psi_{\ell m}-\bar{\Psi}_{\ell m})\sin\th\left[{}_1Y_{\ell m}(\theta,\varphi)+{}_{-1}Y_{\ell m}(\theta,\varphi)\right],\\
F^r_{4\;\ell m}=&-\frac{1}{8}f_0(u^t)^2\mathsf{m}r_0^4\Omega i\sqrt{(\ell -1)(\ell +2)}  \left[\partial^2_t-2f_0\partial_t\partial_r+f_0^2\partial^2_r-\frac{3}{r_0}(1+f_0)\partial_t +\frac{2f_0}{r_0^2}(3r_0-2M)\partial_r \right.\nonumber \\
 &\left.+\frac{2}{r_0^2}(1+2f_0)\right] (\Psi_{\ell m}-\bar{\Psi}_{\ell m})
\sin\th{}_1Y_{\ell m}(\theta,\varphi),\\
F^r_{5\;\ell m}=&-\frac{1}{16}f_0(u^t)^2M\mathsf{m}(\ell -1)(\ell +2)\left(r_0\partial_t -r_0f_0\partial_r -2\right)(\Psi_{\ell m}+\bar{\Psi}_{\ell m})\sin^2\th\,{}_2Y_{\ell m}(\theta,\varphi),\\
F^r_{6\;\ell m}=&\frac{1}{16}f_0(u^t)^2M\mathsf{m}\left[r_0^3\partial^2_t\partial_r -2r_0^3f_0\partial_t\partial^2_r +r_0^3f_0^2\partial^3_r +6r_0^2\partial_t^2  -2r_0(9r_0-13M)\partial_t\partial_r \right. \nonumber \\
 &\left. +12r_0^2f_0(r_0-M)\partial_r^2-6(5r_0-4M)\partial_t+\frac{2}{r_0} (17r_0^2-32r_0M+8M^2)\partial_r  \right. \nonumber \\
 &\left. -\frac{16}{r_0^2}\left(M^2-r_0^2\right)\right](\Psi_{\ell m}+\bar{\Psi}_{\ell m})\sin^2 \th\, {}_2Y_{\ell m}(\theta,\varphi).
\end{align}
\end{subequations}

 Using the definitions of $\eth$ and $\bar\eth$ in terms of partial derivatives with respect of the angular coordinates we can express the spin-weighted spherical harmonics :
\begin{align}\label{sYtoY}
\sqrt{\ell (\ell+1)}\,{}_{1}Y_{\ell m}(\theta,\varphi)=&-\left(\partial_\th - m\csc \th \right)Y_{\ell m}(\theta,\varphi), \\
-\sqrt{\ell (\ell+1)}\,{}_{-1}Y_{\ell m}(\theta,\varphi)=&-\left(\partial_\th + m\csc \th \right)Y_{\ell m}(\theta,\varphi),\\
\sqrt{(\ell-1)\ell (\ell+1)(\ell+2)}\,{}_{2}Y_{\ell m}(\theta,\varphi)=&\left(\partial^2_\th-\cot \th \partial_\th+2m\cot\th\csc \th -2 m\csc \th \partial_\th +m^2 \csc^2\th\right)Y_{\ell m}(\theta,\varphi), \\
\sqrt{(\ell-1)\ell (\ell+1)(\ell+2)}\,{}_{-2}Y_{\ell m}(\theta,\varphi)=&\left(\partial^2_\th-\cot\th\partial_\th-2m\cot\th\csc\th +2 m\csc\th \partial_\th +m^2 \csc^2 \th\right)Y_{\ell m}(\theta,\varphi),
\end{align}
where we have used $\partial_\varphi Y_{\ell m}(\theta,\varphi)\equiv im Y_{\ell m}(\theta,\varphi)$, and $Y_{\ell m}(\theta,\varphi)$ are the usual scalar spherical harmonics. Writing the angular functions in this way allows us to use the same formulas as \citep{BaSa} to re-expand Eqs.\ \eqref{Fr16IRG} and \eqref{Fr16ORG} in spherical harmonics.

As a sum of a single spherical harmonic we get
\begin{equation}
\label{ForceYlm}
F^r_{\ell m}=Y_{\ell m}(\theta,\varphi)\left\{ {\cal F}^r_{(-2) \ell -2, m}+{\cal F}^r_{(-1) \ell -1, m}+{\cal F}^r_{(0) \ell m}+{\cal F}^r_{(+1) \ell +1, m}+{\cal F}^r_{(+2) \ell +2, m} \right\},
\end{equation}
where
\begin{align}
{\cal F}^r_{(-2) \ell m}=& \alpha_{(-2)}^{\ell m}f^r_{2\,\ell m}+(f^r_{5\,\ell m}+f^r_{6\,\ell m})\frac{(-\beta_{(-2)}^{\ell m}+\gamma_{(-2)}^{\ell m})}{\sqrt{(\ell -1)\ell(\ell +1)(\ell +2)}}\pm \frac{\beta_{(-2)}^{\ell m}}{\sqrt{\ell (\ell +1)}} f^r_{4\,\ell m}, \nonumber \\
{\cal F}^r_{(-1) \ell m}=& \pm\frac{\delta_{(-1)}^{\ell m}}{\sqrt{\ell(\ell +1)}}f^r_{4\,\ell m} +\frac{2m\epsilon_{(-1)}^{\ell m}}{\sqrt{(\ell -1)\ell (\ell +1)(\ell +2)}}f^r_{6\,\ell m}, \nonumber \\
{\cal F}^r_{(0) \ell m} =& f^r_{1\,\ell m}+f^r_{2\, \ell m}\alpha_{(0)}^{\ell m}+(\mp f^r_{4\,\ell m}+2f^r_{3\,\ell m})\frac{m}{\sqrt{\ell(\ell +1)}}+(f^r_{5\,\ell m}+f^r_{6\,\ell m})\frac{(-\beta_{0}^{\ell m}+\gamma_{(0)}^{\ell m}+m^2)}{\sqrt{(\ell -1)\ell (\ell +1)(\ell +2)}}, \nonumber \\
{\cal F}^r_{(+1) \ell m}=& \pm\frac{\delta_{(+1)}^{\ell m}}{\sqrt{\ell(\ell +1)}}f^r_{4\,\ell m} +\frac{2m\epsilon_{(+1)}^{\ell m}}{\sqrt{(\ell +1)\ell (\ell +1)(\ell +2)}}f^r_{6\,\ell m}, \nonumber \\
{\cal F}^r_{(+2) \ell m}=&\alpha_{(+2)}^{\ell m}f^r_{2\,\ell m}+(f^r_{5\,\ell m}+f^r_{6\,\ell m})\frac{(-\beta_{(+2)}^{\ell m}+\gamma_{(+2)}^{\ell m})}{\sqrt{(\ell -1)\ell(\ell +1)(\ell +2)}}\pm\frac{\beta_{(+2)}^{\ell m}}{\sqrt{\ell (\ell +1)}} f^r_{4\,\ell m},
\end{align}
the functions $f^r_{i\,\ell m}$ correspond to the angle-independent coefficient of Eqs.\ \eqref{Fr16IRG} or \eqref{Fr16ORG}. Notice the sign dependence of the coefficient multiplying $f^r_{4\, \ell m}$ --- the upper sign is for the IRG modes while the lower sign for the ORG modes. The coupling coefficients $\alpha^{\ell m}$, $\beta^{\ell m}$, $\gamma^{\ell m}$, $\delta^{\ell m}$ and $\epsilon^{\ell m}$ are given explicitly in \citep{BaSa}.

\begin{table}
    \centering
    \begin{tabular}{|c|l|l|c|}
    	\hline $r_0/M$ &  $F^{r\,{\rm Num}}(r_0) \times \frac{M^2}{\mathsf{m}^2}$ & $F^{r\,{\rm MST}}(r_0)\times \frac{M^2}{\mathsf{m}^2} $ \\  
\hline   
  6   & 0.03350126(1)     & 0.033501265(1) \\ \hline
 7   & 0.026070691(5)    & 0.0260706936(1) \\ \hline
 8   & 0.020941671(3)    & 0.02094167456(7) \\ \hline
 9   & 0.017214435(1)    & 0.01721443676(8) \\ \hline
 10  & 0.0144093850(9)   & 0.01440938542(6) \\ \hline
 12  & 0.0105299277(5)   & 0.01052992732(2) \\ \hline
 14  & 0.008031952(1)    & 0.00803195180(1) \\ \hline
 16  & 0.006328227(1)    & 0.006328226988(6) \\ \hline
 18  & 0.005114225(1)    & 0.005114225196(3) \\ \hline
 20  & 0.0042187145(9)   & 0.004218713944(1) \\ \hline
 24  & 0.003011654(1)    & 0.0030116542558(6) \\ \hline
 28  & 0.002257118(5)    & 0.0022571178017(2) \\ \hline
 32  & 0.001754261(4)    & 0.0017542618884(1) \\ \hline
 36  & 0.001402452(3)    & 0.00140245195919(6) \\ \hline
 40  & 0.0011467454(5)   & 0.00114674532583(3) \\ \hline
 50  & 0.0007465337(2)   & 0.00074653378046(1) \\ \hline
 60  & 0.00052437948(8)  & 0.000524379436446(3) \\ \hline
 70  & 0.00038842358(5)  & 0.000388423560775(1) \\ \hline
 80  & 0.00029922175(3)  & 0.0002992217373675(7) \\ \hline
 90  & 0.00023755802(2)  & 0.0002375580134958(4) \\ \hline
 100 & 0.00019316231(2)  & 0.0001931623007419(2) \\ \hline
 120 & 0.00013491660(1)  & 0.00013491660149634(8) \\ \hline
 140 & 0.000099532396(7) & 0.00009953239215925(3) \\ \hline
 160 & 0.000076441055(5) & 0.00007644105294526(1) \\ \hline
 180 & 0.000060543785(4) & 0.00006054378560513(1) \\ \hline
 200 & 0.000049135297(3) & 0.000049135296208105(1) \\ \hline

    \end{tabular}\caption{Comparison between the radial component of the GSF, for different values of $r_0/M$. The second column corresponds to the values computed using numerical integration of Sasaki-Nakamura equation while the values in the third column are calculated in the ORG using the MST method. The quantities in parenthesis correspond to the estimated error on the last quoted decimal shown. The error in the second column is estimated by changing the numerical parameters of the fitting that contributes to the {\it tail}. The error quoted in the third column is estimated from moving the inner boundary when numerically solving the $\ell =1$, $m =1$ multipole.}\label{rvalues}
\end{table}

\bibliography{biblio}

\end{document}